\documentclass[a4paper,twocolumn,showpacs,nofootinbib]{revtex4}
\usepackage{amsmath}
\usepackage{amssymb}
\usepackage{natbib}
\usepackage{graphicx}
\newcommand{\ENDPROOF}{\ensuremath{\square}}
\newcommand{\PROB}{\ensuremath{\mathfrak{P}}}
\newcommand{\FAM}{\ensuremath{\mathfrak{F}}}
\newcommand{\HILB}{\ensuremath{\mathcal{H}}}
\newcommand{\HHILB}{\ensuremath{\tilde{\mathcal{H}}}}
\newcommand{\BE}{\begin{equation}}
\newcommand{\EE}{\end{equation}}

\usepackage{theorem}
\theoremstyle{break} 
\theoremstyle{break} \newtheorem{lem}{Lemma}
\theoremstyle{break} \newtheorem{Def}{Definition}
\theoremstyle{break} 
\theorembodyfont{\rmfamily}
\theoremheaderfont{\scshape}

\begin{document}
\title{Branch dependence in the ``consistent
  histories'' approach to quantum mechanics}
\author{Thomas \surname{M\"{u}ller}}
%\author{Thomas M\"uller}
\email{Thomas.Mueller@uni-bonn.de}
\affiliation{Institut f\"ur Philosophie, Lenn\'estr.~39, 
 53113 Bonn, Germany}
\date{12 November 2006}
\begin{abstract}
  In the consistent histories formalism one specifies a family of histories as
  an exhaustive set of pairwise exclusive descriptions of the dynamics of a
  quantum system. We define \emph{branching families} of
  histories, which strike a middle ground between the two available
  mathematically precise definitions of families of histories, viz., product
  families and Isham's history projector operator formalism. The former are
  too narrow for applications, and the latter's generality comes at a certain
  cost, barring an intuitive reading of the ``histories''.  Branching families
  retain the intuitiveness of product families, they allow for the
  interpretation of a history's weight as a probability, and they allow one to
  distinguish two kinds of coarse-graining, leading to reconsidering
  the motivation for the consistency condition.
\end{abstract}
\maketitle

\section{Introduction}

The consistent histories approach to quantum mechanics
\cite{griffiths84,gellmann90,gellmann93,omnes94,dowker_kent96,kent2000,griffiths2003}
studies the dynamics of closed quantum systems as a stochastic process
within a framework of alternative possible histories. Such a
framework, or \emph{family of histories}, must consist of pairwise
exclusive and jointly exhaustive descriptions of the system's
dynamics.

This intuitive characterization does not yet state what a family of
histories is mathematically. There are two formal definitions in the
literature: So-called \emph{product families} are straightforward
generalizations of one-time descriptions of a system's properties in
terms of projectors to the case of multiple times \cite{griffiths84}.
The so-called \emph{history projector operator} formalism, introduced
by Isham \cite{isham94}, is vastly more general. However, histories in
that formalism do not necessarily have an intuitive interpretation in
terms of temporal sequences of one-time descriptions.

In our paper we make formally precise the notion of a \emph{branching} family
of histories, which is meant to balance generality and intuitiveness:
histories in such a family do correspond to temporal sequences of one-time
descriptions, yet branching families are much more general than product
families. In the context of quantum histories, the notion of branching, or
branch dependence, was originally proposed by Gell-Mann and Hartle
\cite{gellmann90}. It is invoked informally in many publications, but a formal
definition is so far lacking. Our definition of \emph{branching families of
  histories} is based on the theory of branching temporal logic. Apart from
providing a precise reading of a useful concept, our approach allows us to
comment on the relation between the consistency condition for families of
histories and probability measures on such families. It turns out that the
consistency condition is best viewed not as a precondition for introducing
probabilities, as some authors suggest, but as the requirement that
interference effects be absent from the description of a system's dynamics.
Our definition allows for consistent as well as inconsistent families of
histories. It is therefore neutral with respect to the discussion about the
pros and cons of consistency \cite{dowker_kent96,griffiths98,kent2000}, and we
refrain from taking a stance in that discussion.

Our paper is structured as follows: In section~\ref{sec:conshist}, we
introduce some basic facts about consistent histories and
probabilities and review the definition of product families of
histories and the history projection operator approach. We also
sketch the intuitive motivation for a notion of branch-dependent
families. In our central section~\ref{sec:branching}, we give our formal
definition of branch-dependent families of histories and
prove some relevant properties of such families. In the final
section~\ref{sec:discussion}, we discuss the relation between our
new definition and the two mentioned definitions of families of
histories, and we comment on the consistency condition.

\section{Consistent histories}
\label{sec:conshist}

\subsection{Histories, chain operators, and weights}

In the consistent histories approach, a \emph{history} is specified
via properties of the system in question at a finite number of times.%
\footnote{Continuous extensions of the theory have also been studied
  \cite{isham_etal98}, but these will not be considered in this
  paper.---This section closely follows the notational conventions of
  \cite{griffiths2003}, which book provides a detailed and readable
  introduction to consistent histories.}
The system's properties are expressed through orthogonal projectors%
\footnote{More generally, one can specify POVMs or completely positive
  maps; cf.\ \cite{peres2000a}. We will only consider
projectors in this paper.}
on (closed) subspaces of the system's Hilbert space \HILB, i.e.,
operators $P$ for which
\begin{equation}
  \label{eq:projector}
  P\cdot P = P^\dag = P.
\end{equation}
Thus, a single history $Y^\alpha$ consists of a number of projectors
$P^i_\alpha$ at given times $t_i$, $i=1,\ldots,n$: 
\BE\label{eq:nonrelhist} 
Y^\alpha =
P_\alpha^1 \odot P_\alpha^2 \odot \ldots \odot P_\alpha^n.  
\EE 
So far, the symbol ``$\odot$'' should be read as ``and then''; in
Isham's \emph{history projection operator} version of the history
formalism, the symbol can be read as a tensor product (cf.\ 
section~\ref{sec:hpo}).  

A \emph{family of histories} $\FAM$
(sometimes also called a \emph{framework}) is an exhaustive set of
alternative histories.  In line with most of the literature on
consistent histories, we will only consider finite families in this
paper.

Associated with a history $Y^\alpha$ is a \emph{chain operator}
$K(Y^\alpha)$, which is formed by multiplying together the projectors
$P_\alpha^i$ associated with the times $t_i$ ($i=1,\ldots,n$),
interleaved with the respective unitary time development operators
$T(t_i,t_{i+1})$.  Employing the convention of \cite{griffiths2003},
we define
\BE
K^\dag(Y^\alpha) = P_\alpha^1\cdot T(t_1,t_2)\cdot P_\alpha^2\cdot
\ldots \cdot T(t_{n-1},t_n)\cdot P_\alpha^n.  
\EE
These chain operators are often taken to be \emph{representations} of the
respective histories. This is appropriate in that $K(Y^\alpha)$
correctly describes the successive action of the projectors forming
$Y^\alpha$ on the systen. However, the representation relation is in
general many-one, which may be seen as a disadvantage in that one
cannot recover a history from the associated chain operator uniquely.

The system's dynamics explicitly enters the definition of the chain
operators through the time development operators. Thus, a history
$Y^\alpha$ can have a zero chain operator even though the history
involves no zero projectors; such histories are \emph{dynamically
  impossible}, i.e., ruled out by the system's dynamics.

We assume that the initial state of the system is described by a
density matrix $\rho$, which might be proportional to unity if no
information is given.%
\footnote{Many definitions implicitly take $\rho=I$ in the
  case of lacking information, which will lead to wrong scaling of
  inner products and thus, of weights, by a factor of dim($\HILB$).}
The inner product of two operators, $\langle K_1,K_2\rangle_\rho$,
given $\rho$, is defined via
\BE
\langle K_1,K_2\rangle_\rho = Tr[ \rho\cdot K_1^\dag\cdot K_2 ].
\EE
The chain operators allow us to associate with any history $Y^\alpha$
a \emph{weight} $W(Y^\alpha)$, which is the inner product of the
history's chain operator with itself:
\BE
W(Y^\alpha) = \langle K(Y^\alpha),K(Y^\alpha)\rangle_\rho.
\EE
In general, one would hope that these weights correspond to
probabilities for histories from a given family.  We will comment on
that issue below, but first we review a few notions from probability
theory.

\subsection{Probabilities}
\label{sec:probabilities}

A probability space is a triple $\PROB = \langle S,A,\mu\rangle$,
where $S$ is the sample space (the set of alternatives), $A$ is a
Boolean $\sigma$-algebra on $S$, and $\mu$ is a normalized, countably
additive measure on $A$, i.e., a function
\BE
\mu: A\rightarrow [0,1]\quad\quad
\mbox{s.t.}\quad
\mu(S) = 1,
\EE
and such that for any countable family $(a_j)_{j\in J}$ of disjoint
elements of $A$, $\mu$ is additive:
\BE\label{eq:additivity_of_measure}
\mu\left(\bigcup_{j\in J} a_j\right) = \sum_{j\in J} \mu(a_j).
\EE
For a finite set $S$, the algebra $A$ is isomorphic to the so-called
power set algebra, i.e., the Boolean algebra of subsets of $S$, with
minimal element $\emptyset$ and maximal element $S$; the operations of
join, meet, and complement are set-theoretic union, intersection, and
set-theoretic complement, respectively. In the finite case, $\mu$ is
uniquely specified by its value on the singletons (atoms), and
normalization is expressed by the condition
\BE\label{eq:sumatoms}
\sum_{s\in S} \mu(\{s\}) = 1.
\EE
For our treatment of quantum histories, which follows the literature
in assuming finite families, these latter, simplified conditions are
sufficient: A finite probability space is completely specified by
giving a finite set $S$ of alternatives and an assignment $\mu$ of
nonnegative numbers fulfilling (\ref{eq:sumatoms}). The algebra $A$ is
then given as the power set algebra of $S$, and $\mu$ is extended to
all of $A$ via (\ref{eq:additivity_of_measure}).

\subsection{Families of histories: Product families}
\label{sec:productfamilies}

Intuitively, a family $\FAM$ of histories should consist of an
exhaustive set of exclusive alternatives. Thus any one history
$h\in\FAM$ should rule out all other histories from $\FAM$, and $\FAM$
should have available enough histories to describe any possible dynamic
evolution of the system in question. Exclusiveness must in some way be
linked to orthogonality of projectors. This rules out taking
$\FAM$ to be the set of all time-ordered sets of projectors of the
form (\ref{eq:nonrelhist})---such a family would be far too large. The
simplest way to ensure the requirements of exclusiveness and
exhaustiveness is to fix a sequence of time points
\BE
t_1<t_2<\ldots<t_n
\EE
and to specify, for each time $t_i$, a decomposition
of the identity operator $I$ on the
system's Hilbert space \HILB\ into $n_i$ orthogonal projectors
$\{P^i_1,\ldots,P^i_{n_i}\}$, so that
\BE
P^i_j\cdot P^i_{j'} = \delta_{jj'}\,P^i_j,\quad\quad
\sum_{j=1}^{n_i} P^i_j = I.
\EE
In this case, the index $\alpha$ specifying a history $Y^\alpha$ can
be taken to be the list of numbers $(\alpha_1,\ldots,\alpha_n)$, where
$1\leq\alpha_i\leq n_i$. The size $|\FAM|$ of the family $\FAM$ is
given by
\BE
|\FAM| = n_1\times n_2\times \cdots\times n_n.
\EE
Histories in $\FAM$ are pairwise exclusive, since any two different
histories use different, orthogonal projectors at some time $t_i$.
Such a family is also exhaustive, as at every time, the decomposition
of the identity specifies an exhaustive set of alternatives. Formally,
$\FAM$ corresponds to a cartesian product of decompositions of the
identity at different times.
Product families are thus the obvious generalization of one-time 
descriptions of a system's properties in terms of projectors
to the case of multiple times.

\subsection{Branch-dependent families of histories}
\label{sec:branchingintuitive}

While the construction of a product family of histories is the
simplest way to ensure exclusiveness and exhaustiveness, that
construction is by no means the only possibility. Many authors have
noted that in applications, it will often be necessary to choose a
time point $t_{i+1}$, or the decomposition of $I$ at $t_{i+1}$,
dependent on the projector $P^i_\alpha$ employed at time $t_i$. Thus,
e.g., in order to describe a ``delayed choice'' quantum correlation
experiment \cite{aspect82}, one chooses the direction of spin
projection at time $t_2$ depending on the outcome of a previous
selection event at time $t_1$.

It is intuitively quite clear what such ``branch dependence'' would
mean; eqs.\ (\ref{eq:branch_no_prod1})--(\ref{eq:branch_no_prod4})
below give an example of a branch-dependent family of histories.
However, no formally rigorous description of such families of
histories is available so far.  Before we go on to give such a description in
section~\ref{sec:branching}, we introduce Isham's \cite{isham94}
generalized definition of families of histories in terms of history
projection operators (HPOs), with which our definition of branch
dependence will be compared below.

\subsection{Isham's history projection operators (HPO)}
\label{sec:hpo}

The guiding idea of the history projection operator framework is to
single out, as for product families, $n$ times $t_1,\ldots,t_n$ at
which the system's properties will be described. One then forms the
$n$-fold tensor product of the system's Hilbert space:
\BE
\HHILB = \HILB \otimes \ldots \otimes \HILB
\quad\quad
\mbox{($n$ times)}.
\EE
In that large
history Hilbert space $\HHILB$, a history $Y^\alpha$ is read
as a tensor product of projectors:
\BE
Y^\alpha =
P_\alpha^1 \odot P_\alpha^2 \odot \ldots \odot P_\alpha^n =
P_\alpha^1 \otimes P_\alpha^2 \otimes \ldots \otimes P_\alpha^n.  
\EE
That tensor product operator $Y^\alpha$ is itself a projection
operator on $\HHILB$, a so-called \emph{history projection operator},
fulfilling
\BE
Y^\alpha\cdot Y^\alpha=(Y^\alpha)^\dag=Y^\alpha.
\EE
Along these lines one can give an abstract definition of a family of
histories $\FAM = \{Y^1,\ldots,Y^n\}$ as a decomposition of the
history Hilbert space identity $\tilde{I}$:
\BE
Y^\alpha Y^\beta=\delta_{\alpha\beta}Y^\alpha,\quad
\sum_{Y^\alpha\in\FAM}Y^\alpha = \tilde{I}.
\EE
This generalization is formally rigorous, and it allows for further
(e.g., continuous) extensions. However, the generality comes at a
certain cost, since there is no condition that would ensure that a
history projector $Y^\alpha$ should factor into a product of
$n$ projectors on the system's Hilbert space at the $n$ given times, as in
(\ref{eq:nonrelhist}). History projectors that do factor in this way are
called \emph{homogeneous histories}.
%\cite[p.~2162]{isham94}
As the main motivation for introducing histories is given in terms of
homogeneous histories, some authors have expressed doubts as to
whether the full generality of HPO is really
appropriate \cite[p.~118]{griffiths2003}.

One possibility for constructing a narrower framework is to restrict
the HPO formalism to the homogeneous case by requiring that all the
$Y^\alpha\in\FAM$ be products of projectors of the form
(\ref{eq:nonrelhist}). Such a restriction may be implicitly at work in
\cite{griffiths2003}. However, such a restriction is to some extent
alien to the HPO formalism. Nor does it single out a useful class of
families of histories: as we will show below
(eqs.~(\ref{eq:hom_no_branch1})--(\ref{eq:hom_no_branch4})), not all
homogeneous families are branch-dependent families, and some
homogeneous families do not admit an interpretation of weights in
terms of probabilities.

We now describe an alternative approach in which branch dependence is
the natural result of an inductive definition. Furthermore, the
framework allows one to distinguish two different types of
coarse--graining for families of histories.

\section{A formal framework for branch-dependent histories}
\label{sec:branching}

The idea of a branching family of histories is based on the theory
of branching temporal logic that originated in the
work of Prior \cite{prior67}.%
\footnote{Branching temporal logic has already found many applications in
  computer science, linguistics, and philosophy. The interested reader
  is referred to \cite{belnap2001} for an overview.}

\subsection{Branching structures}

For the purpose of constructing finite families of branching
histories, branching temporal logic boils down to the following
inductive definition of a \emph{branching structure}, which is a set $M$ of
\emph{moments} $m_i$ partially ordered by $\preceq$:%
\footnote{One should think of $\preceq$ in analogy to ``less than or
  equal'' ($\leq$), so there is a companion strict order (excluding
  equality), which is denoted by $\prec$. However, note that $\preceq$
  is only a partial order, i.e., some elements of $M$ may be
  incomparable.}

\begin{Def}[Branching structure]
  (i) A singleton set $M=\{m_0\}$ together with the relation
  $m_0\preceq m_0$ is a branching structure.
  (ii) Let $\langle M,\preceq\rangle$ be a branching structure and
  $m\in M$ a maximal element, and let $m^*_1,\ldots,m^*_n$ be new
  elements.  Let $\preceq^*$ be the reflexive and transitive closure
  of the relation $\preceq$ together with the new relations
  $m\preceq^* m^*_1$, \ldots, $m\preceq^* m^*_n$.  Then the set $M\cup
  \{m^*_1,\ldots,m^*_n\}$ together with the relation $\preceq^*$ is
  again a branching structure.
\end{Def}

\begin{figure}[h]
  \begin{center}
    \includegraphics[width=0.6\linewidth]{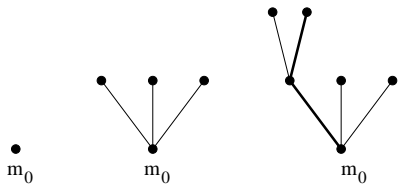}
  \end{center}
\caption{\label{fig:bs}%
Three stages in the construction of a branching structure.
The thick line in the structure on the right indicates one of
the four histories in that structure.}
\end{figure}

By taking a finite number of steps along this definition, one
constructs a finite branching tree in the form of a partially ordered
set with the unique root element $m_0$.%
\footnote{Note that the construction ensures the following formal
  features: The ordering $\preceq$ is transitive (if $x\preceq y$ and
  $y\preceq z$, then $x\preceq z$), reflexive ($x\preceq x$) and
  antisymmetric (if $x\preceq y$ and $y\preceq x$, then $x=y$).
  Furthermore, the ordering fulfills the axioms of ``no backward
  branching'' (if $x\preceq z$ and $y\preceq z$, then either $x\preceq
  y$ or $y\preceq x$) and of ``historical connection'', meaning the
  existence of a common lower bound for any two elements (for any
  $x,y\in M$, there is $z\in M$ s.t.\ $z\preceq x$ and $z\preceq
  y$).---In a more general approach to branching temporal logic, these
  formal features are taken as axioms of a logical framework.}
The maximal nodes in the tree are called ``leaves''.
Figure~\ref{fig:bs} illustrates the inductive process.  Except for the
root element $m_0$, each node has a unique direct predecessor, and except for
the leaves, each node $m$ has one or more direct successors, which
correspond to branching at $m$.  Paths in the tree, i.e., maximal
linearly ordered subsets, extend from the root to one of the leaves
and are thus in one-to-one correspondence with the leaves. These paths
are often called \emph{histories} by logicians, and they can indeed be
given an interpretation in terms of quantum histories, as we will show
in the next section.

\subsection{Branching families of histories}

A branching family of histories can be viewed as a quantum-mechanical
interpretation of a branching structure $\langle M,\preceq\rangle$.
We assume that a system with Hilbert space $\HILB$ (identity operator
$I$) is given.  The interpretation is given by two functions $\tau$
and $P$ that associate times and projectors with the elements of $M$,
respectively.  Formally, we define:
\begin{Def}[Branching family of histories]\label{def:branchingfamily}
  A \emph{branching family of histories} is a quadruple
\BE
 \FAM = \langle M,\preceq,\tau,P\rangle,
\EE 
where \mbox{$\langle M,\preceq\rangle$}
  is a finite branching structure and $\tau$ is a function from $M$ to the
  real numbers respecting the partial ordering $\preceq$:
\BE\label{eq:respectordering}
\mbox{if $m\prec m'$, then $\tau(m)< \tau(m')$}.
\EE
$P$ is a function from $M$ to the set of projectors on $\HILB$ that
assigns projection operators to the elements of $M$ in the following
way: If $m\in M$ is not a maximal element, and $m_1,\ldots,m_{n_m}$
are the $n_m$ immediate successors of $m$, then a set of orthogonal
projectors $P^m_1,\ldots,P^m_{n_m}$ forming a decomposition of the
identity,
\BE\label{eq:associate_projectors}
P^m_i P^m_j=\delta_{ij} P^m_i,\quad
\sum_{i=1}^{n_m} P^m_i = I,
\EE
is assigned to the $m_1,\ldots,m_{n_m}$ via $P(m_i)=P^m_i$.
\end{Def}

The number $\tau(m)\in \mathbb{R}$ is the time associated with $m\in
M$.  While the same time may be assigned to moments in different
histories, we require that $\tau$ respect the partial ordering, as
expressed via (\ref{eq:respectordering}).%
\footnote{The time parameter will only be needed in specifying the
  time development operators in the definition of the chain operators
  later on. Working in the Heisenberg picture, the function $\tau$ would
  not be needed; the condition on $\tau$ would be replaced by
  requiring an appropriate temporal ordering of the Heisenberg
  operators specified through $P$. This already points towards a
  relativistic generalization of the current approach to quantum
  histories, which is currently under preparation \cite{muller_qmbst}.}
As regards the assignment of projection operators, note that
$P(m_i)=P^m_i$ means that at $m$ (the unique predecessor of $m_i$, not
$m_i$ itself), the system had the property expressed by $P^m_i$.%
\footnote{The slightly awkward reference to the previous node in
  (\ref{eq:associate_projectors}) can be avoided if one assigns
  projectors more properly not to nodes, but to \emph{elementary transitions}
  in the branching structure; cf.\ \cite{xu97,belnap2005} for
  details. We stick to our simplified exposition in order not to
  clutter this paper with technicalities---which will, however, be
  relevant for an extension to infinite structures, or to a
  relativistic version employing \emph{branching space-times}
  \cite{belnap92,muller2005,muller_qmbst}.}
The function $P$ thus associates decompositions of the Hilbert space
identity with instances of branching. In this way, each maximal path
$\alpha$ of length $n_\alpha+1$ in $\langle M,\preceq\rangle$,
\BE
m^\alpha_0 \prec m^\alpha_1 \prec \cdots \prec m^\alpha_{n_\alpha},
\EE
corresponds to the $n_\alpha$ elements long chain of projection
operators
\BE
P(m^\alpha_1) \odot P(m^\alpha_2) \odot \cdots \odot P(m^\alpha_{n_\alpha}).
\EE
Here, $m^\alpha_0=m_0$ is the root node, and $m^\alpha_{n_\alpha}$ is
one of the maximal elements. The projectors give information for the
$n_\alpha$ many times
\BE
\tau(m^\alpha_0) < \tau(m^\alpha_1) < \cdots < \tau(m^\alpha_{n_\alpha-1}).
\EE

\begin{figure}[h]
  \begin{center}
    \includegraphics[width=0.6\linewidth]{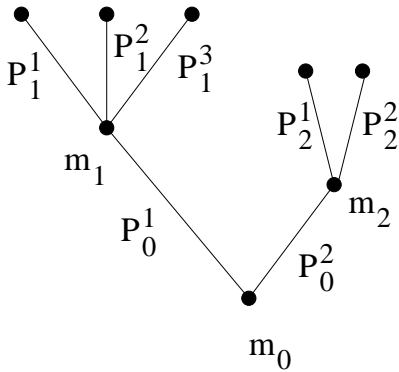}
  \end{center}
\caption{\label{fig:branchingfamily}%
An example branching family of histories. See text for details.}
\end{figure}

Our Definition~\ref{def:branchingfamily} captures the intuitive notion of
branch dependence described in section~\ref{sec:branchingintuitive} in a
formally exact manner, as illustrated by Figure~\ref{fig:branchingfamily}. In
that example of a branching family of histories, the root node $m_0$ has two
successors, $m_1$ and $m_2$, corresponding to the system's having the
properties corresponding to projection operators $P_0^1$ and $P_0^2$ at
$\tau(m_0)$, respectively. The vertical position of $m_1$ vs.\ $m_2$ indicates
that $\tau(m_1)\neq \tau(m_2)$, which is one aspect of branch dependence that
is not available in product families: the time for which the system's property
is described after $\tau(m_0)$ depends on the system's property at
$\tau(m_0)$.  Furthermore, the decompositions of the Hilbert space identity at
$m_1$ ($\{P_1^1,P_1^2,P_1^3\}$) and at $m_2$ ($\{P_2^1,P_2^2\}$) are
different, thus exhibiting the second form of branch dependence that is
not available in product families.

\subsection{Properties of quantum branching histories}

We first note that every product family of histories (cf.\ 
section~\ref{sec:productfamilies}) is a branching family of a very
symmetric kind.

\begin{lem}[Branching vs.\ product families]\label{lem:branching_product}
Every product family of histories is a branching family of histories,
but not conversely.
\end{lem}

\noindent
\emph{Proof:} In order to see that product families are branching
families, let a product family corresponding to $n$ decompositions of
the identity $\{P^i_1,\ldots,P^i_{n_i}\}$ at times $t_1,\ldots,t_n$ be
given. An equivalent branching family is constructed in $n+1$ steps as
follows: We start with a single node, $M_0 = \{m_0\}$ (stage $0$).
Then at each stage $i$ ($1\leq i\leq n$), we add new nodes and enlarge
the structure. We assign the time $t_i$ to all of the maximal elements
of $M_{i-1}$, and we add the decomposition
$\{P^i_1,\ldots,P^i_{n_i}\}$ above all these maxima by introducing $n_i$ new
elements above \emph{each} maximum, thus arriving at the new set of
nodes $M_i$. When $M_n$ has been constructed, we finally assign some
time $t^*>t_n$ to all the maximal elements in $M_n$. This construction
yields a symmetrically growing tree that in the end (at stage $n$)
corresponds to the original product family of histories.

For a branching family that is not a product family, let $\HILB$ have
dimension~2, and let $\{|\phi_1\rangle,|\phi_2\rangle\}$ 
and $\{|\psi_1\rangle,|\psi_2\rangle\}$ be
two different orthonormal bases of $\HILB$. The family
\begin{eqnarray}
\label{eq:branch_no_prod1}
h_1 &=& |\phi_1\rangle\langle\phi_1| \odot |\phi_1\rangle\langle\phi_1|\\
\label{eq:branch_no_prod2}
h_2 &=& |\phi_1\rangle\langle\phi_1| \odot |\phi_2\rangle\langle\phi_2|\\
\label{eq:branch_no_prod3}
h_3 &=& |\phi_2\rangle\langle\phi_2| \odot |\psi_1\rangle\langle\psi_1|\\
\label{eq:branch_no_prod4}
h_4 &=& |\phi_2\rangle\langle\phi_2| \odot |\psi_2\rangle\langle\psi_2|
\end{eqnarray}
describes the system at two times $t_1$ and $t_2$, yielding
a branching family, but not a product family.
\hfill\ENDPROOF
\medskip

Branching families of histories $\FAM = \langle M,\preceq,\tau,P\rangle$
thus yield a natural generalization of product histories while
retaining a strong link between the formalism and the intended
temporal interpretation of the histories. Furthermore, the inductive
definition of the structure makes it easy to prove two key properties
of branching families of histories. Firstly, the construction
immediately shows that $\FAM$ is exclusive and exhaustive: The
one-element family has that property, and it is retained by adding
inductively further (exclusive and exhaustive) decompositions of the
identity at a maximal node.%
\footnote{``Exhaustive'' does not mean ``maximally detailed''. The
  latter notion is quite dubious anyway, as for every finite
  description of a quantum system's dynamics one can give a more
  detailed one.}
Secondly, one can show that the weights of histories in a branching
family $\FAM$ always add up to one. Thus, the weights immediately
induce probabilities on the power set Boolean algebra of $\FAM$ (cf.\ 
section~\ref{sec:probabilities}).

\begin{lem}[Weights in branching families]
\label{lem:addtoone}
  In a branching family of histories $\FAM$, the weights sum to one:
\BE
\sum_{h\in\FAM} W(h) = 1.
\EE
\end{lem}

\noindent
\emph{Proof:}
Assume that an initial state of the system is given by a density
matrix $\rho_0$.%
\footnote{In the case of complete ignorance, $\rho_0 = I /
  \mbox{dim}(\HILB)$.}
For the trivial family consisting of only one history $\{m_0\}$, the
sum of weights reduces to
\BE
W(\{m_0\}) = Tr [\rho_0] = 1.
\EE
Now assume that the property in question holds for the family $\FAM$
corresponding to $\langle M,\preceq,\tau,P\rangle$, let $m$ be a
maximal node, $m^-$ its (unique) direct predecessor, and let
\BE
P^m_i P^m_j=\delta_{ij} P^m_i,\quad
\sum_{i=1}^n P^m_i = I,
\EE
be the decomposition of the identity that is to be employed at the $n$
new maximal elements $m^*_1,\ldots,m^*_n$ that are to be added after
$m$.  Suppose that $\FAM$ had $N$ elements $h_1,\ldots,h_N$, whose
weights add to one. In order to facilitate book-keeping, suppose
further that $m$ is the final node of $h_N$.  To the new quantum branching
structure $\langle M',\preceq',\tau',P'\rangle$ there corresponds a
family $\FAM'$ of $N+n-1$ histories, where for $i=1,\ldots,N-1$,
$h'_i=h_i$, whereas $h_N$ is replaced by the $n$ new histories
$h'_N,\ldots, h'_{N+n-1}$ ending in the new elements
$m^*_1,\ldots,m^*_n$. In order to show that in $\FAM'$, the weights
still add to one, we only need to show that
\BE
W(h_N) = \sum_{i=1}^{n} W(h'_{N+i-1}),
\EE
i.e., the histories replacing old $h_N$ must together have the same
weight as $h_N$. Now in terms of the chain operator $K(h_N)$ for $h_N$,
the chain operators for the new histories are
\BE
K(h'_{N+i-1}) = P^m_i\cdot T(\tau(m),\tau(m^-)) \cdot K(h_N).
\EE
The initial density matrix $\rho_0$, evolved along $h_N$, becomes
\BE
\rho_m = K(h_N)\,\rho_0\,K^\dag(h_N),
\EE
and the weight of $h_N$ can be expressed as
\BE
W(h_N) = Tr[K(h_N)\,\rho_0\,K^\dag(h_N)] = Tr[\rho_m].
\EE
The weights for the new histories can then be written as
\begin{eqnarray}
\lefteqn{W(h'_{N+i-1}) = }  \\
\nonumber
& & 
Tr[P^m_i\, T(\tau(m),\tau(m^-))\, 
\rho_m\, T(\tau(m^-),\tau(m))\, (P^m_i)^\dag] = \\
\nonumber
& & 
Tr[T(\tau(m^-),\tau(m))\, P^m_i\, T(\tau(m),\tau(m^-))\, \rho_m],
\end{eqnarray}
where we used the cyclic property of the trace and $P^\dag = P^2 = P$.
Now as $T$ is unitary, the $n$ operators 
\BE 
\tilde{P}^m_i = T(\tau(m^-),\tau(m))\, P^m_i\, T(\tau(m),\tau(m^-)) 
\EE
are again projectors forming a decomposition of the identity, so that
by the linearity of the trace,
\begin{eqnarray}
\sum_{i=1}^n W(h'_{N+i-1}) &=& \sum_{i=1}^n Tr[\tilde{P}^m_i\, \rho_m]\\
&=& Tr\left[\sum_{i=1}^n \tilde{P}^m_i\, \rho_m\right]\\
&=& Tr[I\cdot \rho_m] = W(h_N),
\end{eqnarray}
which was to be proved.
\hfill\ENDPROOF

\section{Coarse graining, probabilities and the consistency condition}
\label{sec:discussion}

The general idea of coarse graining is that it should be possible to
move from a more to a less detailed description of a given system in a
coherent way.  If probabilities are attached to a fine-grained
description, then an obvious requirement is that the probability of a
coarse-grained alternative should be the sum of the probabilities of
the corresponding fine-grained alternatives. Considerations of coarse
graining are important for quantum histories because of the interplay
between weights of histories and probability measures in a family of
histories. Our discussion will show that one needs to distinguish two
notions of coarse graining.

One notion of coarse graining comes for free in any probability
space: Due to the additivity of the measure, if $b^*$ is the disjoint
union of $b_1,\ldots,b_n$ in the event algebra, then $\mu(b^*) =
\sum_{i=1}^n \mu(b_i)$.%
\footnote{In the infinite case, that equation holds for countable
  unions.}
By Lemma~\ref{lem:addtoone}, for any branching family (and thus, by
Lemma~\ref{lem:branching_product}, for any product family) of quantum
histories, the weights $W(h)$ of the histories $h\in\FAM$ induce a
probability measure on $\FAM$ via $\mu(\{h\}) = W(h)$, which is
extended to the power set Boolean algebra of $\FAM$ via
(\ref{eq:additivity_of_measure}):
\BE\label{eq:cg_sets}
\mu(\{h_1,\ldots,h_n\}) = \sum_{i=1}^n \mu(\{h_i\}) = \sum_{i=1}^n W(h_i).
\EE
If $h_1,\ldots,h_n$ are fine-grained descriptions of a system's
dynamics, eq.~(\ref{eq:cg_sets}) shows that the coarse-grained
description $\{h_1,\ldots,h_n\}$ automatically is assigned the correct
probability.  Thus, branching families of histories unconditionally
and naturally support this notion of coarse graining.

If all branching families support probabilities and coarse graining, then what
is behind the consistency condition? In the literature it is often suggested
that the possibility of defining probabilities or the possibility of coarse
graining for a family of histories is conditional upon the so-called
\emph{consistency condition},
\BE\label{eq:consistency}
\langle K(Y^\alpha), K(Y^\beta)\rangle_\rho = 0\quad
\mbox{if $\alpha\neq\beta$},
\EE
which demands that the chain operators of different histories must be
orthogonal. Condition (\ref{eq:consistency}) is also called ``medium
decoherence'' \cite{gellmann93}.

The above considerations show that for branching families, both probabilities
and one notion of coarse graining are unproblematic, independent of any
condition like (\ref{eq:consistency}). However, that condition does play an
important role with respect to a second, different notion of coarse graining.

That second notion of coarse graining is based on the idea of
constructing from the histories $h\in\FAM$ not \emph{sets of
  histories}, as in eq.~(\ref{eq:cg_sets}), but \emph{new histories},
by something like addition. Whether such additive combination of
histories is possible at all, generally depends on what histories are
mathematically. We will see below that additive combination is not
always possible for histories in a branching family. Accordingly, in
order not to suggest that addition of histories is always
unproblematic, we will use the formal notation ``$sum(h_1,h_2)$'' when
we wish to leave open the question whether that sum is in fact
defined.

If we consider the additive combination of two histories, $h_1$ and
$h_2$,
\BE\label{eq:sumanyhist}
h = sum(h_1,h_2),
\EE
the idea behind coarse graining suggests that for $h$, which is a 
less detailed description than the two fine-grained histories,
the probabilities should just add:
\BE\label{eq:cg_hist}
\mu(h) = \mu(sum(h_1,h_2)) = \mu(h_1) + \mu(h_2).
\EE
Even apart from the question of whether $sum(h_1,h_2)$ can be defined,
eq.~(\ref{eq:cg_hist}) is problematic as it stands: No family of
histories can contain both two histories $h_1$ and $h_2$ and their sum
$h$, as that would violate the requirement of exclusiveness. Thus,
$\mu$ in (\ref{eq:cg_hist}) cannot be a probability measure in a
single family of histories. At this point the idea of weights $W(h)$
as probabilities enters. Assuming that $h=sum(h_1,h_2)$ is indeed a
history, $W$ is defined for all three of $h_1$, $h_2$, and $h$, and
the main idea of (\ref{eq:cg_hist}) can be reformulated as
\BE\label{eq:cg_hist_weights}
W(h) = W(sum(h_1,h_2)) = W(h_1) + W(h_2).
\EE
The validity of (\ref{eq:cg_hist_weights}) is indeed linked to the consistency
condition (\ref{eq:consistency}). However, depending on which type of family
of  histories one considers, there are some subtle issues, as the following
sections point out.

\subsection{Coarse graining in product families}

If $\FAM$ is a product family of histories, the sketched idea of
coarse graining makes immediate sense, as the sum of any two histories
in a given product family can be defined. To consider the basic case,
let two histories $h_1,h_2\in\FAM$ be given,
\BE
h_\alpha = P^1_\alpha\odot \cdots \odot P^n_\alpha,\quad\quad
\alpha=1,2,
\EE
such that they coincide everywhere except for the $j$-th position:
$P^i_1 = P^i_2$ for $i\neq j$, $P^j_1\neq P^j_2$. In this case, one
can define their sum
\BE\label{eq:sumprodhist} 
h = sum(h_1,h_2) := P^1_1 \odot \cdots \odot
P^{j-1}_1 \odot (P^j_1+P^j_2) \odot P^{j+1}_1 \odot \cdots \odot
P^n_1, 
\EE
i.e., at each time $t_i$ for which the histories $h_1$ and $h_2$ are
defined, their sum, $h$, specifies either the same projector as each
of $h_1$ and $h_2$, or gives a less detailed description in terms of
the projector $P^j_1+P^j_2$. The assumption of a product family is
crucial in this definition, as it guarantees that the $P^i_1$ and
$P^i_2$ are both defined at the same times, and that $P^j_1$ and
$P^j_2$ commute---for a branch-dependent family, $P^j_1+P^j_2$, even
if defined, wouldn't normally be a projector.

For a history like $h$ in (\ref{eq:sumprodhist}), the weight function
$W$ is naturally defined even though $h\not\in\FAM$, and it appears
natural to demand that
\BE\label{eq:addweights}
W(h) = W(sum(h_1,h_2)) = W(h_1) + W(h_2).
\EE
This equation does not hold in general in product families of
histories. A family of histories $\FAM$ must satisfy the above-mentioned
condition of \emph{consistency} if it is to satisfy
(\ref{eq:addweights}) for all $h_1,h_2\in\FAM$, and it was along these
lines that Griffiths \cite{griffiths84} originally motivated the
consistency condition for product families of histories.%
\footnote{\label{fn:consistency}%
An extended discussion of questions of uniqueness conditions for
  probability assignments in product families is given in
  \cite{nistico99}.---The notion of consistency in \cite{griffiths84},
  which corresponds to (\ref{eq:addweights}), is weaker than the
  consistency condition (\ref{eq:consistency}) formulated above: For
  (\ref{eq:addweights}) to hold, it is sufficient that the \emph{real
    part} of $\langle K(Y^\alpha),K(Y^\beta)\rangle_\rho$ vanish for
  $\alpha\neq\beta$. The latter condition is known as \emph{weak
    consistency}. In what follows, we will not differentiate between
  medium and weak consistency.}
However, as the next section shows, the symmetric nature of product families
hides an important asymmetry in adding histories.

\subsection{Coarse graining in branching families}
\label{sec:2cg}

We have already seen that in product families $\FAM$, the formal
addition $sum(h_1,h_2)$ can be defined for any $h_1,h_2\in\FAM$. For
branching families, this is not always possible. In fact, we will see
that with respect to formula (\ref{eq:addweights}) one should
distinguish two types of coarse graining, which we call
\emph{intra-branch} and \emph{trans-branch} coarse graining.
\emph{Intra-branch} coarse graining means that maximal nodes from an
otherwise shared branch are added, whereas \emph{trans-branch} coarse
graining means adding ``across branches''. The notion of intra-branch
coarse graning and the respective summation of histories for the basic
case can be defined as follows:

\begin{Def}[Intra-branch coarse graining]
\label{def:intrabranch}
In a branching family $\FAM$, the formal summation $h=sum(h_1,h_2)$ of two
histories $h_1,h_2\in\FAM$,
\BE
h_\alpha = P^1_\alpha\odot \cdots \odot P^{n_\alpha}_\alpha,\quad\quad
\alpha=1,2,
\EE
is called \emph{intra-branch coarse graining} iff $n_1=n_2$, the
histories are defined at the same times, and for $1\leq i<n_1$, $P^i_1
= P^i_2$. In that case, the sum is defined to be
\BE\label{eq:intra_cg}
h=sum(h_1,h_2) := P^1_1\odot \cdots \odot P^{n_1-1}_1
\odot (P^{n_1}_1 + P^{n_1}_2).
\EE
\end{Def}

Intra-branch coarse graining is both well-defined and probabilistically
unproblematic for all branching families:

\begin{lem}\label{lem:intrabranch}
  For intra-branch coarse graining $h=sum(h_1,h_2)$ as in
  (\ref{eq:intra_cg}) in a branching family of histories, the weights
  add according to (\ref{eq:addweights}), i.e.,
\[
W(h) = W(sum(h_1,h_2)) = W(h_1) + W(h_2).
\]
\end{lem}

\noindent
\emph{Proof:} We can follow the lines of the inductive proof of
Lemma~\ref{lem:addtoone}: Let two histories $h_1$ and $h_2$ fulfilling
Definition~\ref{def:intrabranch} be given, and assume that $m$ is the
immediate predecessor of maximal nodes $m^*_1$ and $m^*_2$ of histories
$h_1$ and $h_2$, respectively. Let $m^-$ be the unique direct
predecessor of $m$.  Let $K(h_m)$ be the chain operator for the path
from the root node to $m$, and set $T=T(\tau(m),\tau(m^-))$. Then the
chain operators for the histories $h_\alpha$ are:
\BE
K(h_\alpha) = P(m^*_\alpha)\cdot T\cdot K(h_m).
\EE
The sum of the two final projectors, $P(m^*_1)+P(m^*_2)$, is again a
projector in virtue of (\ref{eq:associate_projectors}).  Accordingly,
the weight of the coarse-grained history $h=h_1+h_2$ is
\begin{eqnarray}
\nonumber
W(h) &=& \langle K(h), K(h)\rangle_\rho
      = \langle K(h_1+h_2), K(h_1+h_2)\rangle_\rho\\
\nonumber
&=& Tr [(P(m^*_1)+P(m^*_2))\, T\, K(h_m)\, \rho\, \\
\nonumber
& & \quad\quad K^\dag(h_m)\, T^\dag\,
      (P(m^*_1)+P(m^*_2))^\dag] \\
&=& Tr [ T^\dag\, P(m^*_1)\, T\, K(h_m)\, \rho\, K^\dag(h_m)] + \\
\nonumber
& &    Tr [ T^\dag\, P(m^*_2)\, T\, K(h_m)\, \rho\, K^\dag(h_m)]\\
\nonumber
&=& W(h_1) + W(h_2),
\end{eqnarray}
where we employed $P^\dag(m^*_\alpha) = P(m^*_\alpha) =
(P(m^*_\alpha))^2$, the fact that $P(m^*_1)\cdot P(m^*_2) = 0$
(\ref{eq:associate_projectors}), and linearity and the cyclic property
of the trace.
\hfill\ENDPROOF
\medskip

So all is well probabilistically if histories are formed as sums of
histories that differ only at the last node, i.e., via intra-branch
coarse graining. Note that this result carries over to product
families of histories: for intra-branch coarse graining in branching
families and in product families, eq.~(\ref{eq:addweights}) holds
automatically, without having to presuppose a consistency condition
like (\ref{eq:consistency}).

What about trans-branch coarse graining? For a product family,
eq.~(\ref{eq:sumprodhist})  shows how to build histories from other
histories quite generally, and we have mentioned the fact that the
validity of eq.~(\ref{eq:addweights}) for trans-branch coarse
graining in a product family generally depends on a consistency
condition like (\ref{eq:consistency}) \cite{griffiths84,griffiths2003}.
For the more general case of branching families, so far the summation
of histories for trans-branch coarse graining has not been defined. It
is possible to define that type of summation in the extended framework
of Isham's HPO formalism, but this amounts to discarding the intuitive
idea that histories are temporal sequences of one-time descriptions
 (cf.\ the next subsection).

The problem of defining trans-branch coarse graining while holding on
to the intuitive interpretation of a history may  be illustrated by
considering a two-dimensional Hilbert space and the two histories
$h_1$ and $h_3$ from eq.~(\ref{eq:branch_no_prod1}) and
eq.~(\ref{eq:branch_no_prod3}), respectively, taken to be defined at
the two times $t_1$ and $t_2$. How should one define $sum(h_1,h_3)$?
Surely one can coarse-grain by considering the \emph{set} of histories
$\{h_1,h_3\}$, and the sum of $h_1$ and $h_3$ in Isham's formalism
amounts to this exactly. But no temporal interpretation in terms of a
single history is forthcoming, as the projectors involved at $t_2$ do
not commute.

Thus, for trans-branch coarse graining, the formal summation $sum(h_1,h_2)$
generally must remain undefined. With respect to the coarse-graining criterion
(\ref{eq:addweights}), this means that in all cases in which it generally
makes sense to ask whether it is satisfied, it is satisfied automatically:
Eq.~(\ref{eq:addweights}) is generally defined only for intra-branch coarse
graining, and for that case, Lemma~\ref{lem:intrabranch} has shown the
equation to hold unconditionally.

This result does of course not mean that all branching families of histories
are consistent. The test of eq.~(\ref{eq:consistency}) can still be applied
for any branching family, and many branching families will be classified as
inconsistent. However, in these cases, the link with
eq.~(\ref{eq:addweights}), which holds for product families, can no longer be
made in our branching histories framework. This points to a somewhat different
interpretation of the consistency condition: In a branching family, that
condition should not be read as a precondition for the assignment of
probabilties via weights (which is unproblematic in view of
Lemma~\ref{lem:addtoone}), but rather as the condition that the different
descriptions of the system's dynamics given by the histories in the family
amount to wholly separate, interference-free alternatives. This is just what
it means for he chain operators to be orthogonal, i.e., to satisfy
eq.~(\ref{eq:consistency}). We hold it to be an advantage of the branching
family formalism proposed here that by leaving trans-branch coarse graining
undefined, it forces us to rethink the interpretation of the consistency
condition. The formalism of product histories is deceptively smooth in
treating all times in the same way, thus blurring the distinction between
intra-branch and trans-branch coarse graining.

\subsection{Coarse graining in Isham's HPO}

The comparison between branching families and Isham's HPO scheme is
illuminating in another respect. HPO is more general and more abstract than
branching families. However, that abstractness comes at a price: we will argue
that much of the intuitiveness of branching families is lost by moving to the
HPO scheme.

In Isham's HPO scheme, histories are themselves represented as
projectors on the (large) history Hilbert space, not as chain
operators on the (much smaller) system Hilbert space. As an HPO-family
$\FAM$ must correspond to a decomposition of the history identity
operator, sums of histories in $\FAM$ will again correspond to history
projectors (even though not from the given family); the formal
addition $h=sum(h_1,h_2)$ has a direct interpretation as the literal
addition of history projectors. Thus one can form a Boolean algebra with
elements
\BE\label{eq:hpo_algebra}
Y = \sum_{Y^\alpha\in\FAM} \pi_\alpha Y^\alpha,\quad\quad
\pi_\alpha\in\{0,1\},
\EE
that is isomorphic to the power set algebra of $\FAM$ (the
$\pi_\alpha$ playing the role of characteristic functions)---just like
in the first type of coarse graining considered at the beginning of
this section. From one point of view, addition here always forms like
objects from like objects: sums of history projectors are again
history projectors. From another point of view, however, addition
remains problematic: Even if the $Y^\alpha$ are homogeneous histories,
i.e., have an intuitive interpretation as temporally ordered sequences
of projectors, that will generally not be so for their sums. Thus in
general, the elements (\ref{eq:hpo_algebra}) of the Boolean algebra
are inhomogeneous histories without an intuitive interpretation.

At the level of abstraction of eq.~(\ref{eq:hpo_algebra}), one need not
distinguish between different kinds of coarse graining. Accordingly, it seems
more natural to demand additivity of weights,
\BE\label{eq:addweights_hpo}
W\left(\sum_{Y^\alpha\in\FAM} \pi_\alpha Y^\alpha\right) =
\sum_{Y^\alpha\in\FAM} \pi_\alpha W(Y^\alpha),
\EE
which corresponds to satisfaction of the consistency condition
(\ref{eq:consistency}).  However, at the same level of abstraction, one can
note that $W$ is a quadratic function, whereas (\ref{eq:addweights_hpo})
demands linearity---not a natural demand at all. The motivation for additivity
was, after all, given in terms of the time-ordered sequences of projectors
that formed the basis of the history framework, not in terms of some abstract
algebra. Furthermore, even when HPO is restricted to families of homogeneous
histories, a probability interpretation of the weights is not forthcoming
generally; families of HPO histories can violate a number of seemingly
straightforward assumptions \cite{isham_linden94}. As an example, consider the
following family of histories: Let $\HILB$ have dimension 2, and let
$\{|\phi\rangle,|\psi\rangle\}$ be an orthonormal basis, so that
$\langle\phi|\psi\rangle = 0$. Then, define
\BE
|\chi\rangle = \frac{1}{\sqrt{2}} (|\phi\rangle + |\psi\rangle);\quad\quad
|\chi'\rangle = \frac{1}{\sqrt{2}} (|\phi\rangle - |\psi\rangle).
\EE
Note that $\langle\chi|\chi'\rangle=0$. We now construct a family of
two-time histories; the corresponding history Hilbert space $\HHILB$
has dimension 4:
\begin{eqnarray}
\label{eq:hom_no_branch1}
h_1 &=& |\chi\rangle\langle\chi| \odot |\psi\rangle\langle\psi|\\
h_2 &=& |\chi'\rangle\langle\chi'| \odot |\psi\rangle\langle\psi|\\
h_3 &=& |\phi\rangle\langle\phi| \odot |\phi\rangle\langle\phi|\\
\label{eq:hom_no_branch4}
h_4 &=& |\psi\rangle\langle\psi| \odot |\phi\rangle\langle\phi|
\end{eqnarray}
These four histories are pairwise orthogonal, and their sum is the
identity operator in $\HHILB$. Thus, $\{h_1,\ldots,h_4\}$ is a
homogeneous family of histories. Now, taking the initial density
matrix $\rho$ to be the pure state $\rho=|\phi\rangle\langle\phi|$,
one can compute the following:
\begin{eqnarray}
K(h_1)\,\rho\,K^\dag(h_1) = \frac{1}{4} |\psi\rangle\langle\psi| ;\quad W(h_1) &=& \frac{1}{4}\\
K(h_2)\,\rho\,K^\dag(h_2) = \frac{1}{4} |\psi\rangle\langle\psi| ;\quad W(h_2) &=& \frac{1}{4}\\
K(h_3)\,\rho\,K^\dag(h_3) = |\phi\rangle\langle\phi| ;\quad W(h_3) &=& 1\\
K(h_4)\,\rho\,K^\dag(h_4) = 0 ;\quad W(h_4) &=& 0
\end{eqnarray}
Thus, the sum of weights in this HPO family of histories is $3/2$,
barring any straightforward probability interpretation. This family is
\emph{not} a branching family of histories in the sense of
section~\ref{sec:branching}, proving that branching families of
histories are a subclass even of homogeneous HPO families.%
\footnote{Note that the temporally reversed family \emph{is} a
  branching family, and the weights do sum to unity (as only the
  mirror image of $h_3$, which is $h_3$ itself, contributes a non-zero
  weight).}

\subsection{Discussion}

The consistent history approach to quantum mechanics offers a view of
quantum mechanics that honours many classical intuitions while
remaining, of course, faithful to the empirical predictions of
orthodox quantum mechanics. While the initial motivation of the
approach in terms of product families makes good pedagogical sense, it
is too narrow for applications. Furthermore, as we have shown in
section~\ref{sec:2cg}, product families may be misleading because they
fail to distinguish between two importantly different notions of
coarse-graining.

A number of applications demand branching families of histories.
However, so far there has not been available a formally rigorous
definition of that class of families. Isham's HPO formalism, while
offering the necessary generality, is in danger of losing touch with
the intuitive motivation of the history approach. To be sure, this
does not amount to any fundamental criticism, but it gives additional
support for providing a less general definition that stays closely
tied to the intuitive motivation of branch-dependent histories.

Through our definition of branching families of histories we have here
provided the sought-for formal framework.  Branching families are more
general than product families, and they are general enough for
applications while retaining a natural interpretation.

\begin{figure}[!]
  \begin{center}
    \includegraphics[width=0.6\linewidth]{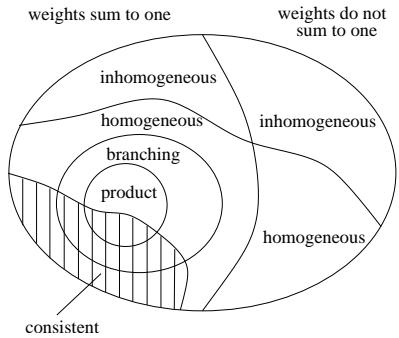}
  \end{center}
\caption{\label{fig:fam}%
The relation of the various notions of families of histories considered
in this paper.}
\end{figure}

Figure~\ref{fig:fam} gives a graphical overview of the various kinds of
families of histories that we considered in this paper.  Branching families
always admit an interpretation of weights in terms of probabilities, as the
weights of the histories in such a family add to one. Consistent branching
families in addition are free from interference effects.  The formal framework
presented here is neutral with respect to the question of whether families of
histories that are not consistent in this sense can be put to good physical
use or not.

\begin{acknowledgments}
  I would like to thank audiences at Oxford and Pittsburgh for stimulating
  discussions. Special thanks to Nuel Belnap, Jeremy Butterfield, Jens Eisert,
  Robert Griffiths, and Tomasz Placek.  Thanks to Przemys\l{}aw Gralewicz and
  to four anonymous reviewers for additional helpful comments.  Support by the
  Polish KBN and by the Alexander von Humboldt-Stiftung is gratefully
  acknowledged.
\end{acknowledgments}

%
%\nocite{*}
% to produce all citations in the database
%

\bibliographystyle{plain}
%\bibliography{qmbst}

\end{document}